\documentclass[12pt]{spieman}  
\usepackage{amsmath,amsfonts,amssymb}
\usepackage{graphicx}
\usepackage{setspace}
\usepackage{tocloft}
\usepackage{color,soul}
\usepackage[dvipsnames]{xcolor}
\usepackage{multicol}

\title{Laboratory Demonstration of the Wrapped Staircase Scalar Vortex Coronagraph}

\author[a*]{Niyati Desai}
\author[b]{Garreth Ruane}
\author[a]{Jorge Llop-Sayson}
\author[a]{Arielle Bertrou-Cantou}
\author[b]{Axel Potier}
\author[b]{A J Eldorado Riggs}
\author[b]{Eugene Serabyn}
\author[a]{Dimitri Mawet}
\affil[a]{California Institute of Technology, Pasadena, California, United States}
\affil[b]{Jet Propulsion Laboratory, California Institute of Technology, Pasadena, California, United States}

\cftpagenumbersoff{figure}
\cftpagenumbersoff{table} 
\begin{document} 
\maketitle

\begin{abstract}

Of the over 5,000 exoplanets that have been detected, only about a dozen have ever been directly imaged. Earth-like exoplanets are on the order of 10 billion times fainter than their host star in visible and near-infrared, requiring a coronagraph instrument to block primary starlight and allow for the imaging of nearby orbiting planets. In the pursuit of direct imaging of exoplanets, Scalar Vortex Coronagraphs (SVCs) are an attractive alternative to Vector Vortex Coronagraphs (VVCs). VVCs have demonstrated $2 \times 10^{-9}$ raw contrast in broadband light but have several limitations due to their polarization properties. SVCs imprint the same phase ramp as VVCs on the incoming light and do not require polarization splitting, but are inherently chromatic. Discretized phase ramp patterns such as a wrapped staircase help reduce SVC chromaticity and simulations show it outperforms a chromatic classical vortex in broadband light. We designed, fabricated and tested a wrapped staircase SVC and here we present the broadband characterization on the High Contrast Spectroscopy Testbed (HCST). We also performed wavefront correction on the In-Air Coronagraph Testbed (IACT) at NASA's Jet Propulsion Laboratory and achieved an average raw contrasts of $3.2\times 10^{-8}$ in monochromatic light and $2.2\times 10^{-7}$ across a 10\% bandwidth.
 
\end{abstract}


{\noindent \footnotesize\textbf{*}Address all correspondence to Niyati Desai,  \linkable{ndesai2@caltech.edu} }

{\noindent \footnotesize{\textit{Paper received by JATIS Jan. 11, 2023; accepted for publication Mar. 24, 2023; published online Apr. 8, 2023.}}}


\section{Introduction}
\label{sec:intro}  

The age old question of if there is life out there in the universe is the driving curiosity behind much of today's astronomical ambitions. Since their discovery only thirty years ago, thousands of new exoplanets have been detected through various techniques, and many more are expected to be found in the next decade. Of the over 5,000 exoplanets that have been detected, only about a dozen have ever been directly imaged.\cite{Akenson2013} Techniques such as transit and radial velocity make up the majority of exoplanet detections that are confirmed today; however, these methods have limitations regarding the study of key molecules in exoplanet atmospheres or surface temperatures over a wide range of planet types and separations from the host star.\cite{Astro2020_Report} On the other hand, direct imaging is a promising method for the spectroscopic detection of biosignatures on Earth-like planets orbiting Sun-like stars. However, direct imaging is very challenging because Earth-like exoplanets around Sun-like stars are on the order of $10^{10}$ times fainter than their host star in visible light and Earth-like exoplanets around M-type stars are on the order of $10^{8}$ fainter. Furthermore, for the purpose of atmospheric spectroscopy, imaging the target over a large spectral bandwidth is necessary.

A coronagraph instrument is a promising solution to the challenge of exoplanet direct imaging because when centered directly on a star, it blocks most of the primary starlight and allows for imaging of nearby orbiting planets. The selection of a coronagraph does not only depend on its ability to suppress starlight, but also on factors such as high off-axis throughput (e.g., from a planet), a small inner working angle, and insensitivity to chromatic errors, tip/tilt and other low-order wavefront aberrations. The Astro2020 Decadal Survey has recognized direct imaging of exoplanets as a valuable research direction for space telescope mission concepts such as the Habitable Exoplanet Explorer (HabEx) and the Large Ultra-violet,
Optical, Infrared (LUVOIR) Surveyor\cite{Astro2020_Report,HabExReport, LUVOIRReport}. The science goals of the Astro2020 Decadal Survey are driving motivators for advancements in coronagraph instruments as the next flagship mission will likely be the very first direct imaging mission dedicated to Earth-like exoplanet imaging and characterization. 


The concept of the vortex coronagraph was first introduced in 2005 \cite{Mawet2005,Foo2005}. Over the last fifteen years, it has been explored as a viable alternative to the classical Lyot coronagraph for high contrast imaging because it allows for higher throughput of planet light along with a smaller inner working angle and lower sensitivity to low-order aberrations. \cite{Swartzlander2005, Mawet2005, Foo2005, Lee2006, Swartzlander2008}. 

The primary mechanism of a vortex coronagraph in performing light suppression involves creating an optical vortex from the incoming wavefront by placing a phase mask in the focal plane, and then blocking the light diffracted by the mask with a slightly undersized aperture (the Lyot stop) downstream in the pupil plane. Deformable mirrors (DM) are also usually placed upstream of the focal plane mask (FPM) to correct any wavefront errors and further reduce the amount of starlight in the region of interest for exoplanet detection.

Vortex coronagraphs come in two flavors: vector vortex coronagraphs (VVCs) and scalar vortex coronagraphs (SVCs). Each offers its own advantages and conversely unique limitations (Section~\ref{sec:coronagraphs}). The primary difference lies in how vector and scalar vortex FPMs differently imprint an optical vortex onto incoming wavefronts.

The development of VVCs has received much more attention than SVCs and they have recently been successfully implemented in several ground based observatories. On-sky demonstration and operation of VVCs at Palomar Observatory\cite{Mawet2009}, Keck Observatory\cite{Serabyn2017,Wang2020}, Subaru Observatory\cite{Kuhn2018}, and the VLT\cite{Wagner2021} have proven the capabilities of vortex coronagraphs while experimental testing and further development at Caltech, JPL and other labs continues to push their potential for high contrast imaging of exoplanets\cite{Serabyn2019,Ruane2020,Ruane2022,Konig2022}.

The earliest literature on SVCs only dates back to Swartzlander et al.~2006\cite{Swartzlander2006} who explored their potential and suggested possible limitations and Lee et al.~2006\cite{Lee2006} who presented the first staircase vortex experimental verification. Later, Swartzlander et al.~2008\cite{Swartzlander2008} reported the first demonstration of an SVC. Ruane et al.~2019\cite{Ruane2019} discusses the theoretical potential that SVCs offer. Finally, Galicher et al.~2020\cite{Galicher2020} showed how phase wrapping of SVC topographies impacts the achromatic limitations discussed in Section~\ref{sec:coronagraphs} and Desai et al.~2021\cite{Desai2021} presents chromatic characterization of the phase wrapped staircase SVC. In this paper we present the first laboratory results of the wrapped staircase SVC, a promising alternative to current vector vortex coronagraph instruments.

\section{Scalar Vortex Coronagraph Advantages and Limitations}
\label{sec:coronagraphs}

As mentioned in Section~\ref{sec:intro}, there are two flavors of vortex coronagraph: scalar and vector. SVCs and VVCs both imprint a phase ramp, $\exp(i l \theta)$ where $l$ is the vortex charge, or the number of phase wraps the light does in one wavelength, onto an incoming wavefront to create an optical vortex, but they differ in the mechanism each uses to do so.

SVCs turn incoming wavefronts into an optical vortex through longitudinal phase delays, whereas VVCs use geometric phase shifts. SVCs vary in either surface height or in index of refraction with respect to the azimuth to imprint the desired phase ramp on the wavefront. Conversely, vector vortex masks are essentially a half wave plate with a spatially varying fast axis, which results in a polarization dependence. They are often manufactured from liquid crystal polymers~\cite{Mawet2009LiquidCrystal,Ganic2002}, subwavelength gratings~\cite{Mawet2005Subwavelength}, photonic crystals~\cite{Murakami2013}, or metamaterials~\cite{Genevet2012}. A VVC with charge $l$ still effectively imprints a spiral phase ramp  onto the incoming light, but creates two conjugated phase ramps $\exp(\pm i l \theta)$ on the orthogonal right and left circular polarization state of the light. 

Because of these split direction phase ramps, current VVC implementations involve introducing several additional polarizing optics, which not only require further precise alignment, but also add to the cost and complexity of the optical setup. This dependence on polarization is the primary disadvantage of VVCs because wavefront control with DMs in a single optical train cannot simultaneously provide high-contrast solutions for orthogonal polarization states and correct for polarization-induced aberrations. One solution to this requires introducing a polarizer and an analyzer at the front and end of the light path, ensuring that only one circular polarization reaches the focal plane and enters the vector vortex mask. Another solution is to duplicate the entire instrument for each of the positive and negative phase-ramped polarizations by way of a polarizing beamsplitter. 


The primary motivation behind exploring SVCs is to circumvent the polarization dependent limitation imposed by vector vortex masks. This is a significant advantage because it means not cutting the total throughput in half from filtering out one polarization.

A scalar vortex mask's primary limitation comes from the fact that a longitudinal phase delay is inherently chromatic in behavior. This chromaticity factor can be seen in Equation \ref{eq:transmission}, which shows the complex transmission function of an SVC and includes the wavelength dependence\cite{Ruane2019}. An ideal vortex coronagraph requires $l$ to be a nonzero, even integer value for the starlight to be completely diffracted outside of the Lyot stop~\cite{Swartzlander2005, Mawet2005, Foo2005, Lee2006}. At the design wavelength $\lambda_{0}$, an SVC creates a perfect optical vortex with a spiral phase ramp described by $\exp(i l \theta)$. 

\begin{equation}
\label{eq:transmission}
    t=\exp \left(i \frac{l \lambda_{0}}{\lambda} \theta\right)
\end{equation}

But for light at any wavelength other than the designed central wavelength $\lambda_{0}$, the charge of the vortex will vary. Essentially as the wavelength is offset from the central design wavelength, the charge $l$ becomes non-integral for $\lambda \ne \lambda_{0}$ and the SVC's ability to perform light suppression decreases--hence it is chromatic. Here we explore limitations and performances of a simple type scalar vortex mask.



\section{Simulations of Mask Topographies}
\label{sec:topsims}
A few strategies have been proposed to achromatize SVCs including combining two optics of different refractive indices into a dual-layer FPM (Swartzlander 2006)\cite{Swartzlander2006}, or cleverly wrapping the phase profile of the phase ramp (Galicher et al. 2020)\cite{Galicher2020}. Wrapping the phase of a vortex, as introduced by Galicher et al. 2020, essentially means introducing phase jumps of 2$\pi$ to allow for a continuous vortex phase ramp with the same charge but spanning a smaller overall phase range. Any other non-2$\pi$ phase jumps might lead to off-axis PSF distortions or attenuations.

Here we focused on methods of varying the FPM surface topography. In order to fully understand the effect that the topography of an SVC has on its coronagraphic performance and to effectively choose one design to fabricate for lab testing, we performed a trade study of several designs in simulations as well as Fourier decomposition analyses. Of the designs shown in Section~\ref{sec:topsims}, we fabricated the wrapped staircase scalar vortex mask, and characterized both its light suppression capability and chromaticity both before and after performing wavefront sensing and control. The wrapped staircase mask presented here incorporates discrete steps along the vortex phase ramp similar to the charge 2 staircase mask in Lee et al. 2006\cite{Lee2006}. However, that staircase design did not implement phase wrapping. Here the charge 6 staircase phase ramp is wrapped six times in the design proposed by Ruane et al. 2019\cite{Ruane2019}. One of the primary motivations is to study how this impacts the broadband performance.

\begin{figure*}
\begin{center}
\includegraphics[width=16cm]{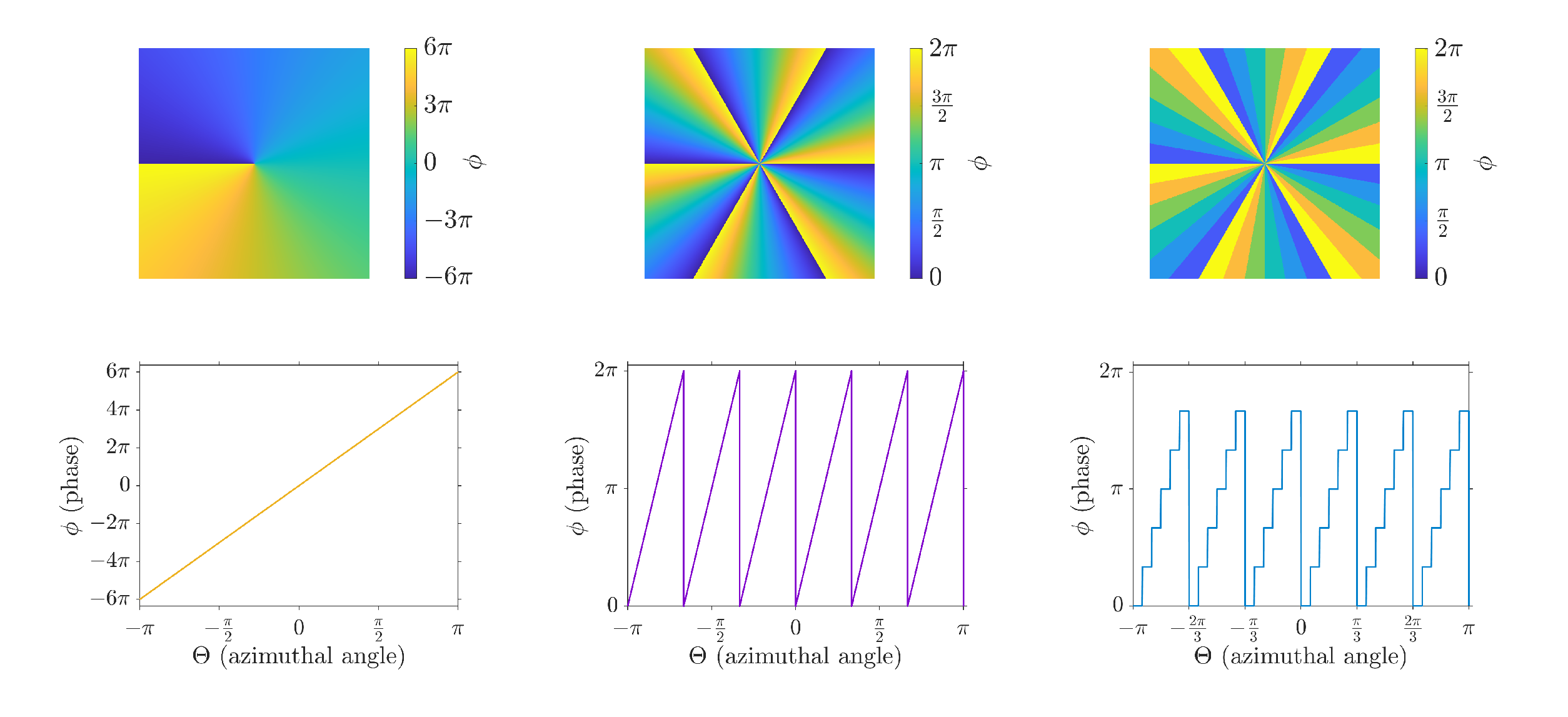}
\end{center}
\caption 
{ \label{fig:phaseprofs}
Phase mappings (top row) and corresponding phase profiles (bottom row) for three charge 6 scalar vortex FPM topographies (from left to right): classical vortex, sawtooth vortex, wrapped staircase vortex. The wrapped staircase vortex (right) was originally proposed in Ruane et al. 2019.\cite{Ruane2019}} 
\end{figure*}

We simulated several different SVC topographies of charge 6 to perform a direct comparison of their performance. Figure~\ref{fig:phaseprofs} shows the phase mappings and the corresponding azimuthal profiles for each of the phase mappings of a charge 6 classical vortex, sawtooth vortex, wrapped staircase vortex. The focus of this paper is to provide a direct comparison for the wrapped staircase scalar vortex phase mask. However, Desai et al. 2022~\cite{Desai2022} considers  the simulated performance of an additional charge 6 variation of the phase wrapping design derived from that of Galicher et al. 2020~\cite{Galicher2020} alongside these three topographies. While the design from Galicher et al. was of a charge 8 coronagraph and the ones considered here are all charge 6, the primary differences between them are the surface topographies and the overall range of the phase ramp spanned by the design. Here we consider three vortex topographies.

\textbf{Classical Vortex:}
A classical vortex FPM is a simple phase ramp wrapping around the optical axis, and its slope is proportional to the charge of the vortex. In Figure~\ref{fig:phaseprofs}, the left image is a charge 6 classical vortex, so its phase is a linear ramp from $-6\pi$ to $+6\pi$. 

\textbf{Sawtooth Vortex:}
In the sawtooth vortex design, which is called this because of its azimuthal profile, the charge of the vortex dictates the number of phase ramps, and therefore the slope. In the middle image of Figure~\ref{fig:phaseprofs},  the charge 6 sawtooth vortex has 6 ramps, each of which starts at a phase of 0 and goes to $2\pi$.

\textbf{Wrapped Staircase Vortex: }
The wrapped staircase vortex design from Ruane et al.~2019\cite{Ruane2019}, whose name is apparent from its phase profile shown in Figure~\ref{fig:phaseprofs}. Instead of several continuous linear phase ramps, this design allows for several sectors with flat steps at heights ranging from 0 to $2\pi$ within each sector. Starting the first step at $\phi = 0$, the staircase does not actually reach 2$\pi$, because of the digitization of the ramp\cite{Ruane2019}. For example, Figure~\ref{fig:phaseprofs} (c) shows a charge 6 wrapped staircase vortex with 6 sectors and 6 steps in each sector.

\subsection{Wavefront Propagation Simulations}
\label{subsec:falcosims}

We used the Fast Linearized Coronagraphic Optimizer (FALCO)\footnote{\url{https://github.com/ajeldorado/falco-matlab}} software package for simulating the point spread function (PSF) of a star, and 2D chromatic wavefront propagation through all subsequent optics including our FPM~\cite{Riggs2018}. To simulate various FPM designs, we provided high resolution phase mappings like the ones shown in Figure~\ref{fig:phaseprofs} and adjusted resolution parameters in FALCO's wavefront propagation to find an appropriate simulation environment for this study. 

FALCO simulation without wavefront control found that on average the wrapped staircase mask attenuates starlight by a factor of approximately $10^{-7}$ across a 10\% bandwidth and approximately $10^{-6}$ across a 20\% bandwidth. The simulated performance of the classical vortex  across a 20\% bandwidth without EFC is a few times $10^{-6}$.

For wavefront control, we simulate electric field conjugation (EFC)\cite{Give'on2007} to iteratively find an optimal DM shape that suppresses residual starlight in the region of interest, called dark hole, where a planet might be found. After 50 iterations of EFC across a 10\% bandwidth, the classical vortex yielded an average contrast of 1.16e-6 whereas the wrapped staircase vortex yielded an average contrast of 3.92e-7.

\subsection{Modal Decompositions of Phase Profiles}
\label{subsec:modaldecomp}
We used a modal decomposition of the phase profiles to get a more extensive understanding of SVC behaviors, as done in Ruane et al. 2019 for the pursuit of understanding the comparative performance of FPM topographies based on their phase profiles\cite{Ruane2019}. We present this analysis tool for predicting the performance of any azimuthally varying SVC topography. It is important to understand why various topographies yield different contrast curves and the modal decomposition tool provides a possible insight in explaining this.

Since an optical vortex is a Fourier series, in the exponential form, each coefficient can be linearly decomposed as shown below.

Any optical vortex can be written as:
\begin{equation}
\label{eq:fourier}
    t(\theta,\lambda)=\Sigma_{m} C_{m}(\lambda)e^{im\theta}
\end{equation}

with

\begin{equation}
\label{eq:coeffs}
    C_{m}(\lambda) = \frac{1}{2\pi} \int_{\pi}^{-\pi} t(\theta,\lambda) e^{-im\theta} d\theta
\end{equation}

And each mode \textit{m} corresponds to a charge m vortex. This can be interpreted to predict the behavior of a new SVC topography as a combination of differently charged vortexes. Modal decomposition is a way to break down and visualize the power distribution of the vortex modes of an SVC, which gives insight into stellar leakage and sensitivity to low order aberrations, as well as the effective topological charge for any azimuthal phase profile.

\begin{figure*}
\begin{center}
\includegraphics[width=16cm]{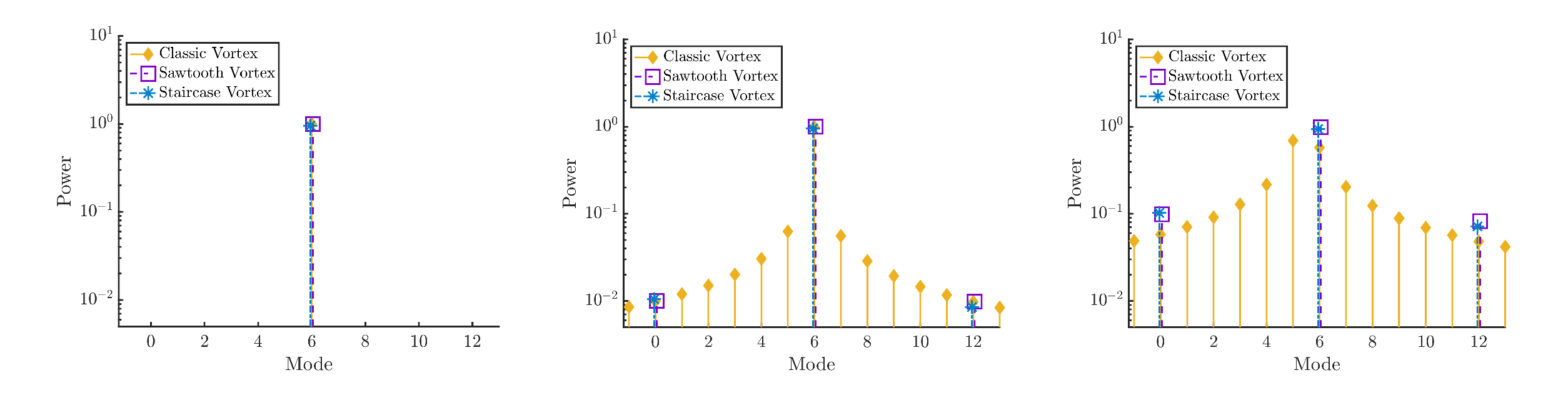}
\end{center}
\caption 
{ \label{fig:moddecomp}
A modal decomposition at a the design wavelength (left), 1\% wavelength offset (center) and a 10\% wavelength offset (right) shows how each of the classical vortex (yellow), sawtooth vortex (purple), and wrapped staircase vortex (blue) topographies differ in behavior for broadband light. } 
\end{figure*}

 We applied this technique to the SVC topographies mentioned in Section~\ref{sec:topsims} to compare their predicted performances. As shown in Figure~\ref{fig:moddecomp} (left), at the design central wavelength, only the expected mode, six, is found and there is consistent behavior across all three FPM topographies. However, when a modal decomposition is performed at a wavelength offset from the central wavelength, the chromatic behavior of each SVC is revealed. In Figure~\ref{fig:moddecomp} at a 1\% offset (middle), and a 10\% offset (right), mode six is still the primary peak, however, a peak emerges at mode zero which corresponds to the zeroth order leakage. Zeroth order leakage corresponds to starlight passing through the coronagraph without being affected by the vortex at all and therefore not being suppressed. It is thus the most unfavorable mode.

An intuitive explanation of the sawtooth design's modal decomposition can be derived from the known modes of a sawtooth wave in signal Fourier analysis. Typically, the first mode is at some $f_{0}$ and all the other modes are at higher frequencies which are multiples of $f_{0}$, harmonics. For a classical SVC (single phase ramp), the first frequency is at the topological charge of 1, then the second one is at 2, etc--explaining the low-order modes in the decomposition for the classical vortex. For a sawtooth design, the first non-zero frequency is $f_{0}=6$, then the next is$f_{0}=12$, etc--matching the results for the sawtooth vortex modal decomposition.

The peaks at even modes correspond to energy distributed to other vortexes with even charge. Therefore, for these modes, starlight is effectively nulled and their behavior in the presence of low order wavefront aberrations is well studied \cite{Ruane2018}. Odd modes, on the other hand, correspond to odd charged vortexes, where starlight suppression is deficient since only an even charge produces a null for a vortex coronagraph~\cite{Swartzlander2005, Mawet2005, Foo2005, Lee2006}. This explains why the classical vortex, which displays odds modes in the modal decomposition (orange), exhibits the worst contrast performance in FALCO simulations~\cite{Desai2022}.

Figure~\ref{fig:moddecomp} suggests that the wrapped staircase and sawtooth topographies outperform the classical vortex, particularly because no other odd or small even modes besides the zeroth order leakage appear in the broadband modal decomposition analysis. Since the modal decompositions for the wrapped staircase and sawtooth vortex topographies are comparable, the wrapped staircase design was chosen to be made for its manufacturability.

\section{Wrapped Staircase Mask Design and Fabrication}

\begin{figure*}
\begin{center}
\includegraphics[height=5.5cm]{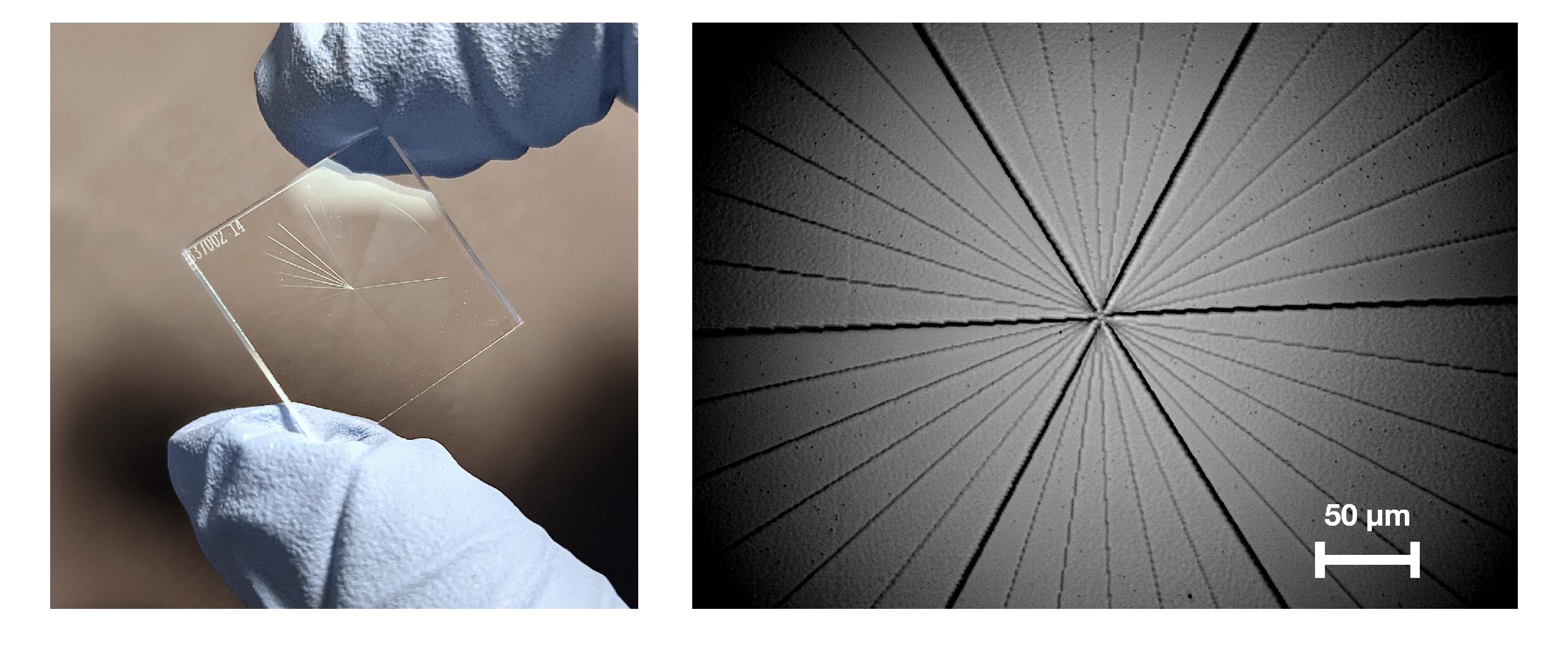}
\end{center}
\caption 
{ \label{fig:svcimage}
A charge 6 wrapped staircase SVC mask fabricated in fused silica(left) and a microscopic image of its central defect (right).} 
\end{figure*}

The scalar vortex mask fabricated was a wrapped staircase scalar vortex mask manufactured by Zeiss and designed to have a charge of 6. It had 6 sectors made up of 6 steps each. The mask was fabricated on fused silica and created for a central wavelength of 775~nm. The square silica sample is 15~mm by 15~mm in size and the actual spiral phase plate region is circular with a diameter of 13mm and centered in the square (see Fig.~\ref{fig:svcimage}).

An important measure of the quality of an FPM, particularly vortex masks, is the size of the central defect because light through the central leakage does not pick up any phase ramp. Microscope measurements revealed the central defect on the mask to be no larger than approximately 8.2 microns in diameter. It is imperative that the central defect be small relative to the beam/spot size. Since the focal ratios of the optical setup here are large (F/\# $\approx$ 30 for HCST and F/\# $\approx$ 80 for IACT), this measured central defect is sufficiently small at the working wavelength of 775 nm. The device has a broadband anti-reflective coating with a reflectance of less than 0.2\% over 725nm to 825nm. Transmittance is taken into account through the normalization since the off-axis PSF image is also taken through the mask.

\section{Chromaticity Characterization}
\label{sect:characterization}
To better understand the chromatic behavior of the wrapped staircase SVC mask, we put the mask in the focal plane of the HCST and looked at its light suppression capabilities at a range of wavelengths.

\subsection{High Contrast Spectroscopy Testbed}

\begin{figure*} [ht]
\begin{center}
\includegraphics[height=5.5cm]{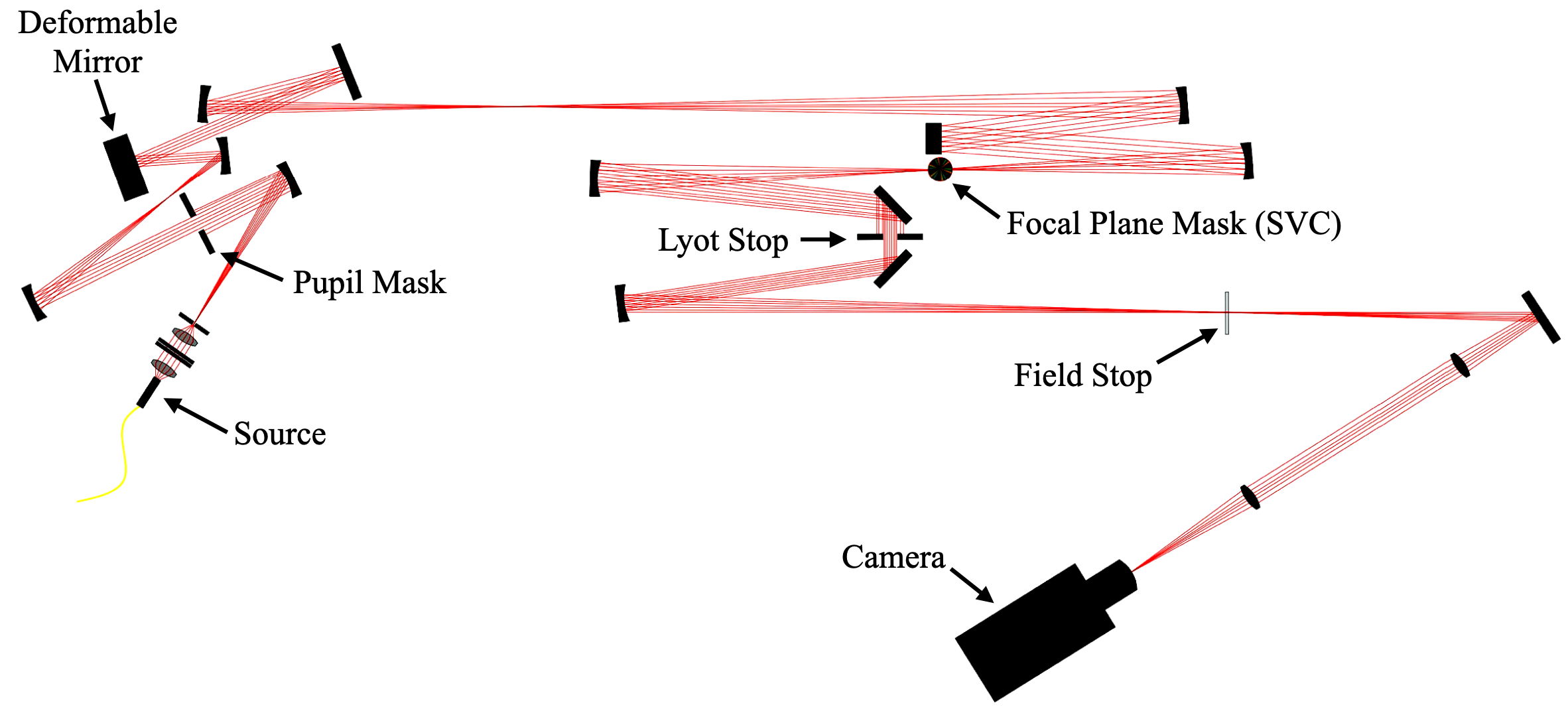}
\end{center}
\caption 
{ \label{fig:hcst} 
Layout of the High Contrast Spectroscopy Testbed for demonstrating the scalar vortex coronagraph. Refer to Llop-Sayson et al. 2020a \cite{Llop-Sayson_SPIE2020} and 2020b\cite{AVC2020} for details about HCST testbed design and setup.}
\end{figure*}

The layout for the HCST is shown in Fig.~\ref{fig:hcst}. A more detailed description of the HCST testbed and the optical components designed and incorporated can be found in Llop-Sayson et al. 2020a \cite{Llop-Sayson_SPIE2020} and 2020b\cite{AVC2020}. The star was simulated using a supercontinuum white-light laser source (NKT Photonics SuperK EXTREME) followed by a tunable single-line filter (NKT Photonics SuperK VARIA) to isolate the specific wavelengths desired for testing. We sampled wavelengths by moving the tunable filter to select narrowbands (6~nm bandwidth) between 685~nm and 820~nm at increments of 15~nm. After the entrance pupil, the adaptive optics system consists of a Boston Micromachines Corporation kilo-DM that flattens the wavefront. The DM has a continuous surface membrane with 34 $\times$ 34 actuators with an inter-actuator separation of 300 $\mu$m. At the coronagraph, the light is focused on the wrapped staircase scalar vortex mask. Next, a circular Lyot stop with $\sim93\%$ of the radius of the 16.4~mm beam blocks the light diffracted by the FPM in the downstream pupil plane. This size of Lyot stop was chosen because a smaller Lyot stop would lower throughput. Lastly, a field stop is introduced to help block any unwanted light outside the region of interest from reaching the camera. The remaining light is imaged with a $\sim$ f/50 beam onto the camera (Oxford Instruments Andor Neo 5.5). A filter wheel is placed before the camera, either set to a neutral density (ND) filter  (Thorlabs NE20B, OD = 2.0) or an empty slot for photometric calibration.

\subsection{Broadband Raw Contrast Results}

\begin{figure*}
\begin{center}
\includegraphics[width=16cm]{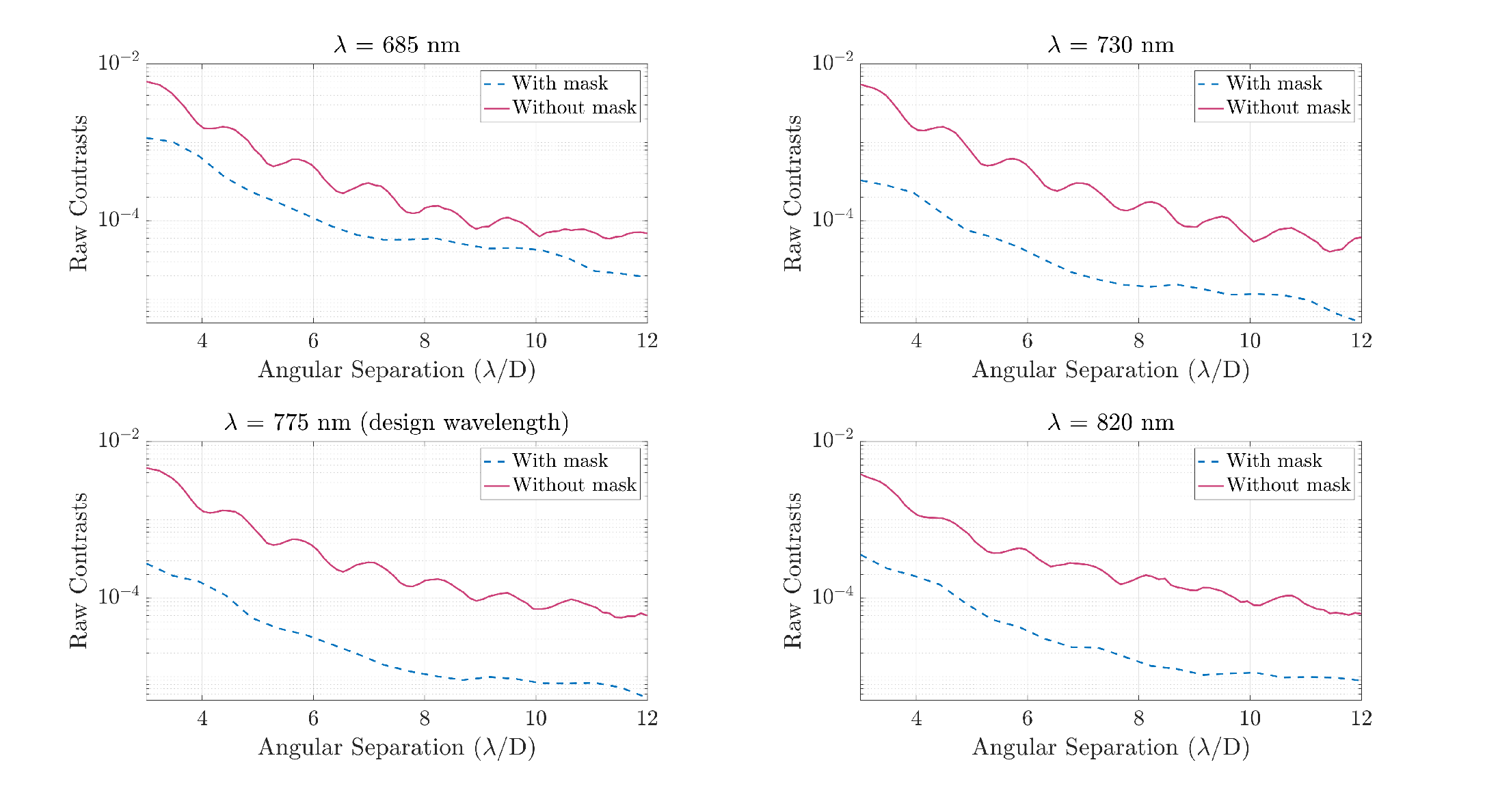}
\end{center}
\caption 
{ \label{fig:contrastprofs}
Experimentally measured unocculted PSF (solid pink) and coronagraphic (dashed blue) contrast curves from focal plane images for 685 nm, 720 nm, 775 nm, and 820 nm, and angular separations from 3 to 12 $\lambda$/D on HCST. Overall, the wrapped staircase scalar vortex coronagraph improved the raw contrasts and suppressed the Airy ring diffraction pattern shown by the bumps in the pink PSF curves. } 
\end{figure*}

To quantitatively characterize the focal plane starlight suppression, the contrast curves in Fig.~\ref{fig:contrastprofs} show the raw contrast at four wavelengths: 685~nm, 730~nm, 775~nm, and 830~nm with angular separations ranging from 3 $\lambda$/D out to 12 $\lambda$/D. Here, raw contrast is defined as the averaged intensity of the coronagraphic image divided by the peak intensity of the unocculted pseudo-star image. The peak intensity is defined as the highest pixel value of the non-coronagraphic image. The error is negligible since the pixel sampling on HCST is approximately 6 pixels per $\lambda/D$ ($3\times$Nyquist). In each plot the pink curve is the contrast profile without the scalar vortex mask; the distinctive bumps correspond to the Airy rings of the diffraction pattern in the PSF. The blue dashed curves are the raw contrasts with the scalar vortex mask. The overall lower contrast and the lack of bumps in the blue dashed curve shows the SVC’s ability to suppress diffraction. Of these four profiles, it is also clear that the largest contrast improvement is with the 775 nm light, which was the design central wavelength for these scalar vortex masks.

\begin{figure*}
\begin{center}
\begin{tabular}{c}
\includegraphics[width=14cm]{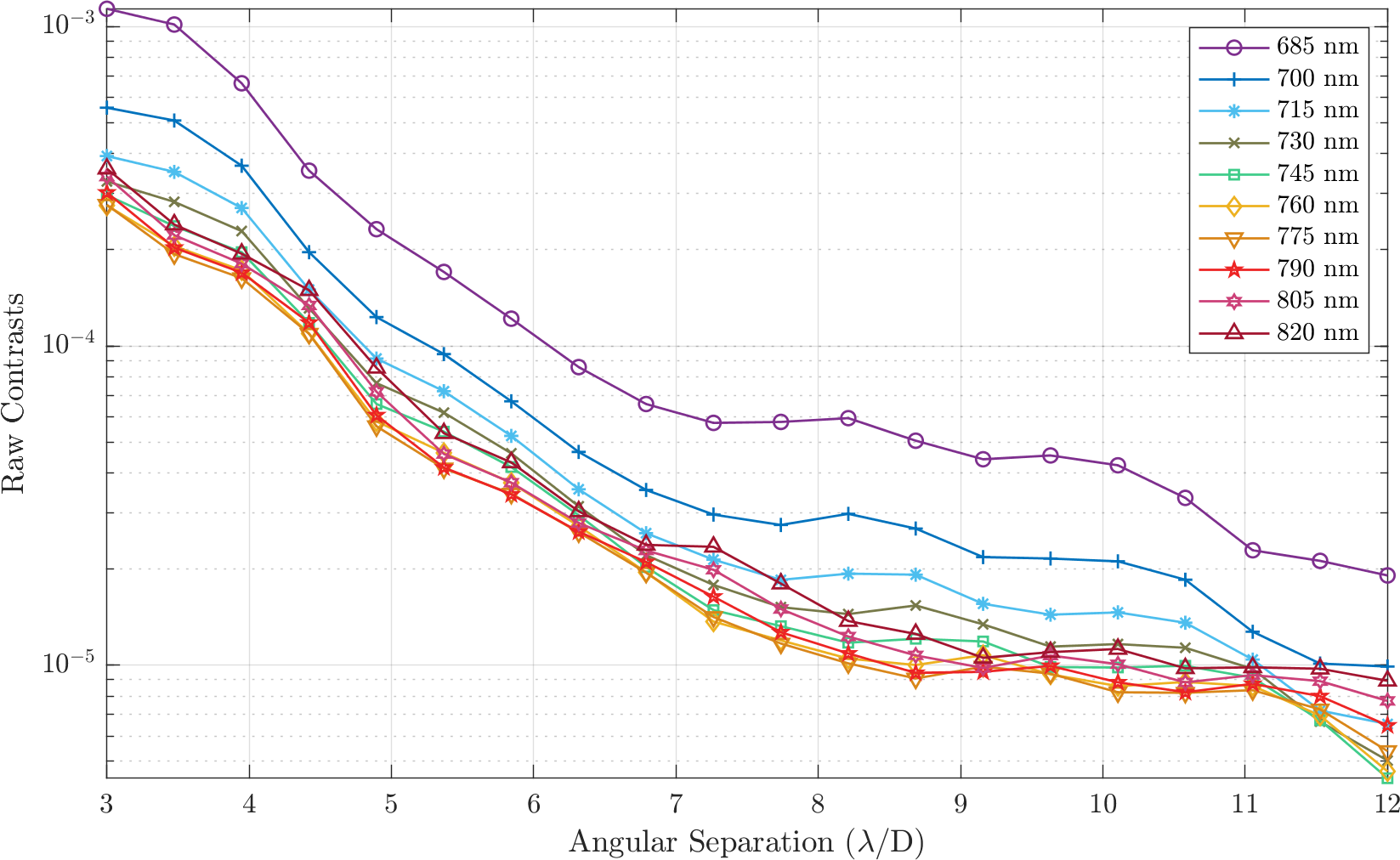}
\end{tabular}
\end{center}
\caption 
{ \label{fig:multicontrasts}
Experimentally measured classical contrast curves for wrapped staircase SVC plotting raw contrast versus angular separation for wavelengths from 685 nm to 820 nm. These were measured on HCST and calculated by taking the ratio of the azimuthal averages at each angular separation to the peak intensity of the unocculted PSF. The best performance is observed at 775~nm (the dark red curve), which was the central design wavelength for this scalar vortex mask.} 
\end{figure*}

By sampling over a large range of wavelengths and taking narrowband measurements at each, we observed the light suppression capability of the SVC was found to degrade as the incoming starlight's wavelength deviated from the central 775 nm, as seen by the green triangle marked curve in Figure~\ref{fig:multicontrasts}.  Furthermore, from this graph, it can be seen that observing light across approximately a 10\% bandwidth (the curve at 760~nm and the curve at 790~nm) degrades the contrast by a factor of 2 or 3.

\begin{figure*}
\begin{center}
\includegraphics[width=16cm]{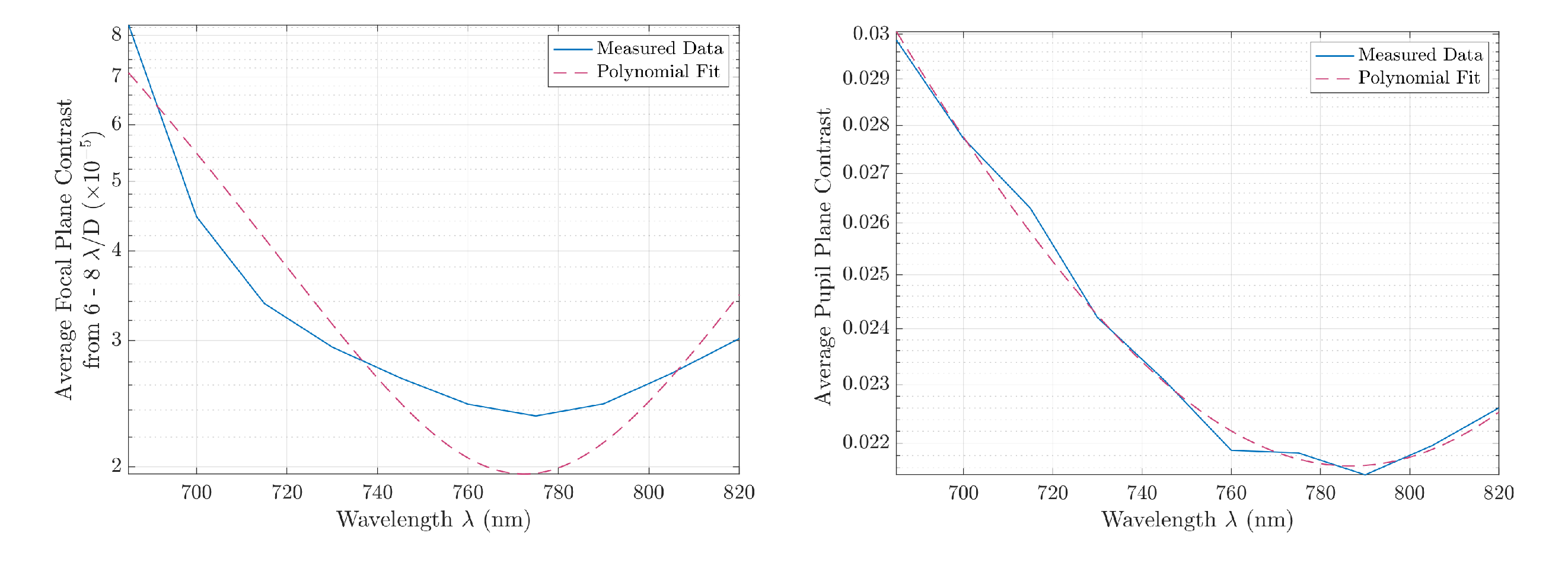}
\end{center}
\caption 
{ \label{fig:focpupcontrasts}
(left) A focal plane raw contrast vs wavelength plot (solid blue) averaged across the region from 6 to 8 $\lambda$/D and a quadratic fit (dashed pink) to that data. (right) A pupil plane raw contrast versus wavelength plot (solid blue), which takes the ratio of the flux inside the pupil to the flux outside the pupil, and quadratic fit (dashed pink).} 
\end{figure*}

The image plane contrast curve as a function of wavelength in Figure~\ref{fig:focpupcontrasts} shows the general chromatic behavior of the SVC, with peak performance close to 775 nm, the design central wavelength for this mask. We consider the region of 6-8 $\lambda/D$ instead of 3-10 $\lambda/D$  for a comparison with the theoretical chromaticity fit in the focal plane because this region farther from the center is not as dominated by low-order wavefront aberrations and would be a reasonable metric for how much the starlight suppression performance changes as a function of wavelength. The pupil plane contrast plot in Figure~\ref{fig:focpupcontrasts} demonstrates in blue the ratio of the total flux inside the pupil vs outside the pupil. In dashed pink the parabolic fit of this curve describes a second order dependence on wavelength, as expected from theory\cite{Ruane2019}. The pupil plane data is expected to be more trustworthy because it integrates across all spatial frequencies and is less limited by the speckle patterns in the focal plane. These experimental results also compare to the simulated FALCO results without wavefront control for monochromatic light at the design wavelength, and at a 10\% offset in Section~\ref{subsec:falcosims}.

\section{Wavefront Sensing and Control Results}
\label{sect:wfsc}

After characterizing the broadband performance of the wrapped staircase scalar vortex mask on HCST, we performed wavefront sensing and control on the In-Air Coronagraph Testbed (IACT) to assess the raw contrast performance of the coronagraph. 

\subsection{In-Air Coronagraph Testbed}

\begin{figure*}
\begin{center}
\begin{tabular}{c}
\includegraphics[width=13cm]{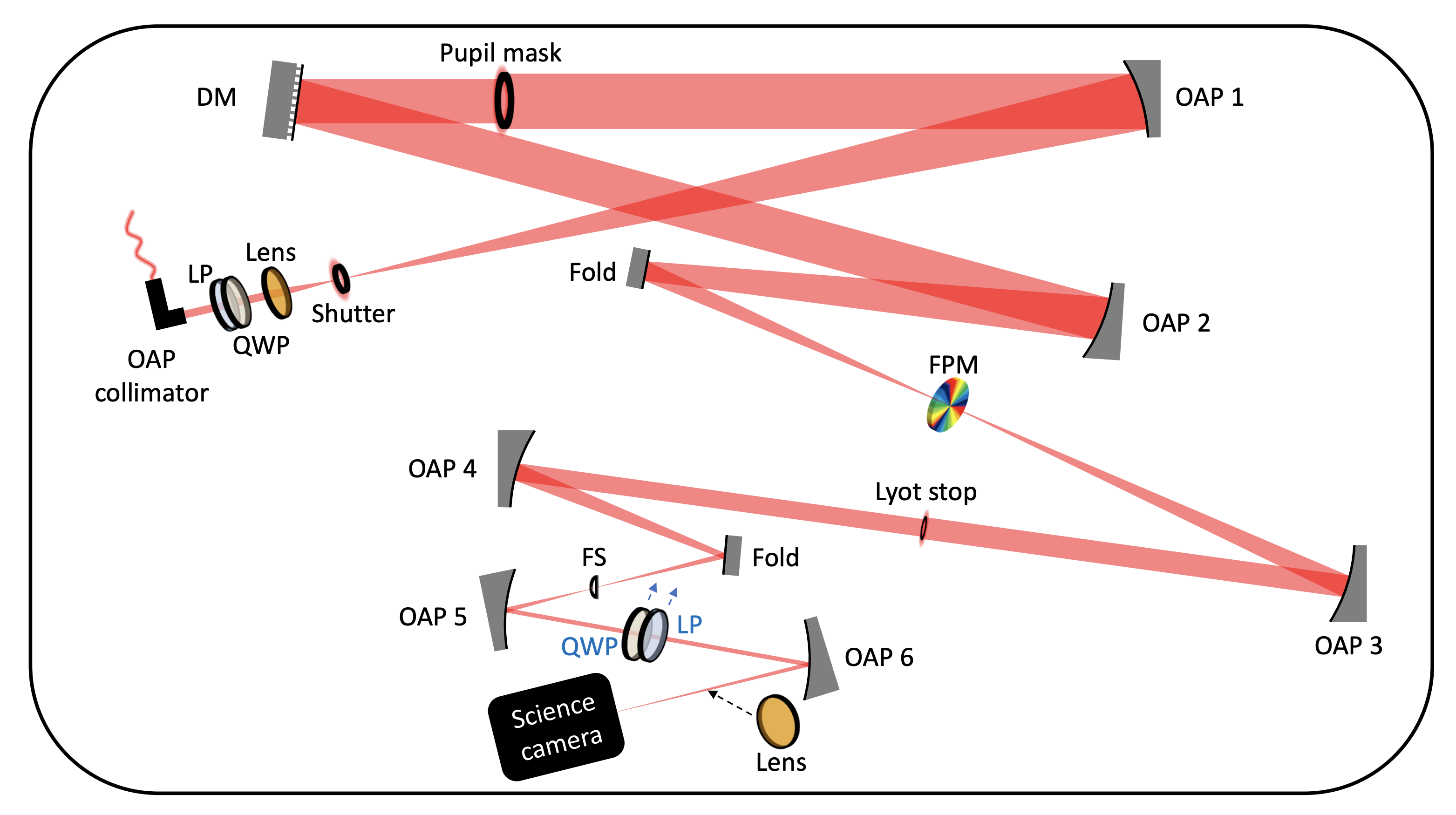}
\end{tabular}
\end{center}
\caption 
{ \label{fig:iact}
Layout of the In-Air Coronagraph Testbed (IACT) for performing wavefront sensing and control the scalar vortex coronagraph. Shown in blue, the quarter wave-plate and linear polarizer before the camera were used with the VVC experiments and moved out of the beam for the SVC experiments. Refer to Baxter et al. 2021 \cite{Baxter2021} for details about IACT testbed design and setup.} 
\end{figure*}

The layout for the IACT is shown in Figure~\ref{fig:iact}. The primary differences between the two in-air testbeds, IACT and HCST, is that IACT has a bigger DM and a larger F number (F/\# $\approx$ 82.5 compared to F/\# $\approx$ 30) at the focal plane mask plane. The DM on IACT is a 2040 actuator (2k) MEMS DM manufactured by Boston Micromachines Corporation (BMC). Here the laser source on IACT was replaced with an NKT for broadband testing. Additionally, to test both the SVC and VVC focal plane masks, the quarter wave-plate and linear polarizer downstream of the coronagraph in Figure~\ref{fig:iact} are used with the VVC and removed with the SVC.

\begin{figure*}
\begin{center}
\includegraphics[width=16cm]{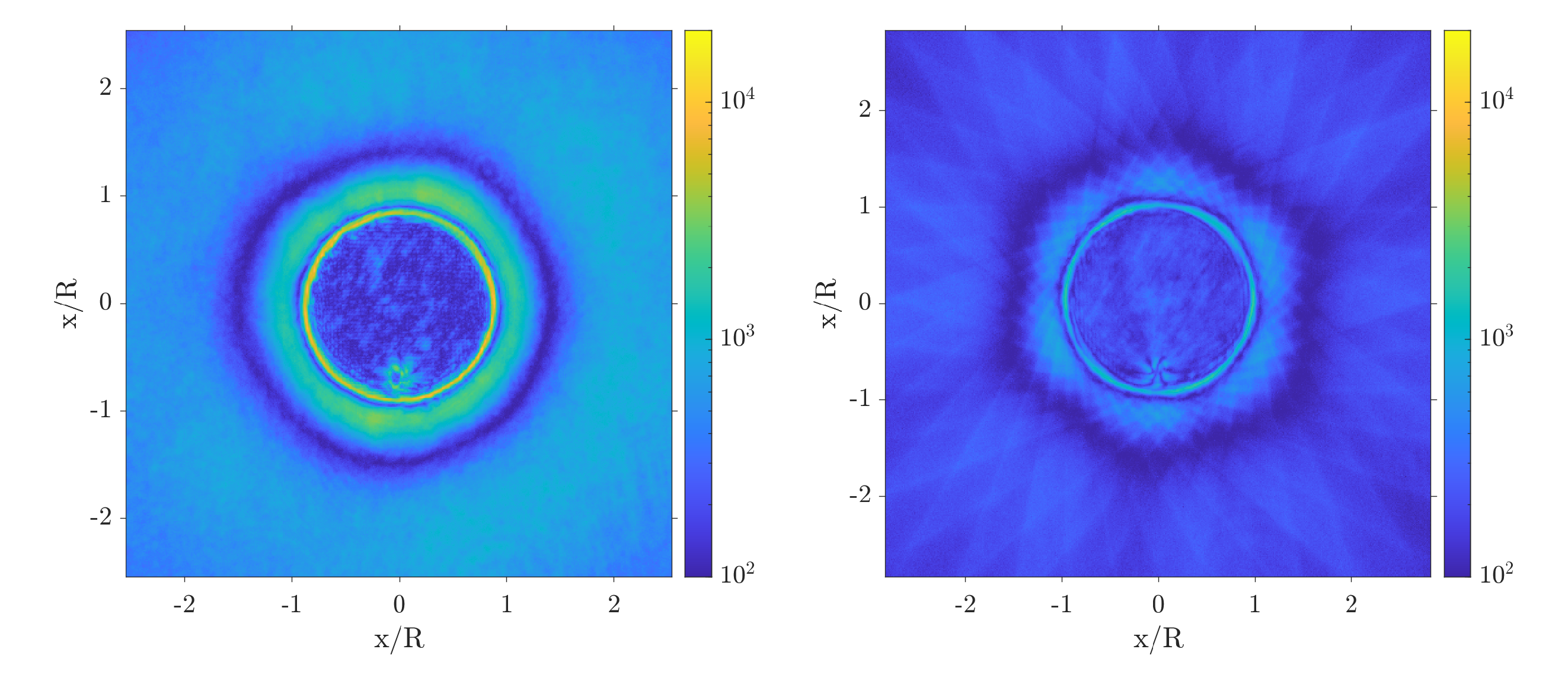}
\end{center}
\caption 
{ \label{fig:pupplanes}
Pupil plane images without the Lyot stop on IACT taken after 0.001 seconds of integration time across a 10\% bandwidth. The ring of fire can be seen for both (left) the vector vortex coronagraph and (right) the scalar vortex coronagraph. Both pupils show a defect in the bottom due to a dead actuator in the deformable mirror. The right pupil image shows the SVC is letting a lot more light through the Lyot stop opening than the VVC.} 
\end{figure*}

Fig.~\ref{fig:pupplanes} shows the pupil plane images taken on IACT with the Lyot stop removed for both a charge 6 VVC manufactured by Beam Co and the SVC previously characterized on HCST. From the images, we can see the distinct “ring of fire” visible along the edges of the pupil plane, which helps to clearly demonstrate how well the FPM is suppressing starlight inside the Lyot stop aperture. The defect visible in both the VVC and SVC pupil plane in the bottom of the pupil comes from a dead actuator in the deformable mirror. The light at the edge of the pupil is a direct indication that the scalar vortex mask is successfully rejecting light. The wrapped staircase SVC pupil plane image here matches closely with that simulated by Ruane et al. 2019\cite{Ruane2019}.

\subsection{Dark Hole Digging}

\begin{figure*}
\begin{center}
\includegraphics[width=16cm]{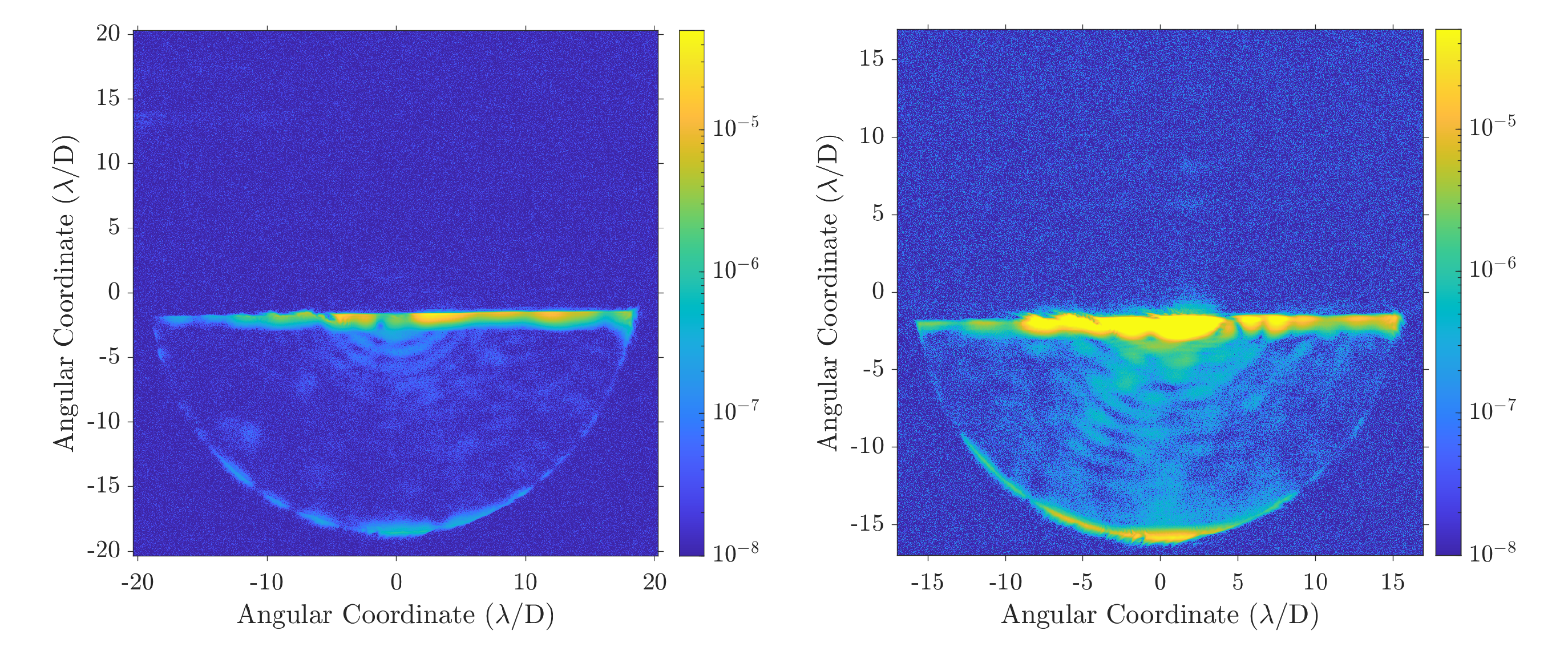}
\end{center}
\caption 
{ \label{fig:darkholes}
The resulting dark holes for the (left) vector vortex coronagraph and (right) scalar vortex coronagraph after performing electric field conjugation on IACT. Both are semicircle dark holes across a 10\% bandwidth. The digging region for the scalar vortex coronagraph is up to 15 $\lambda/D$ and for the vector vortex coronagraph it is up to 20 $\lambda/D$. Across the same scoring region of 3-10 $\lambda/D$, the average contrast is $3.2\times10^{-8}$ for the vector vortex coronagraph's dark hole and $2.2\times10^{-7}$ for the scalar vortex coronagraph's dark hole. } 
\end{figure*}

Measurements on IACT were done for both monochromatic light and broadband light with EFC. Figure~\ref{fig:darkholes} shows the resulting focal plane dark holes after 50 iterations of EFC for the SVC and VVC in broadband side by side for a direct comparison. For the SVC, the dark hole is a semicircle digging region from 3 to 15 $\lambda/D$. For the VVC, the digging region is a semicircle from 3 to 20 $\lambda/D$ since it was measured at a lower central wavelength of 635~nm, compared to 775~nm for the SVC. In monochromatic and broadband, for both the SVC and VVC the scoring region, or the area over which the average contrast is measured is from 3-10 $\lambda/D$.

In monochromatic light, for the SVC, the average contrast across the scoring region, or the area over which the average contrast is measured was $3.2\times10^{-8}$. For the VVC in monochromatic light the average contrast in the same scoring region was $1.2\times10^{-8}$. In broadband, the average contrast across the scoring region for the SVC was $2.2\times10^{-7}$ taken over a 10\% bandwidth. Across the same 10\% bandwidth, the average contrast in the scoring region for the VVC was $3.2\times10^{-8}$.

The wrapped staircase vortex experimental results compare to the expected simulated values from FALCO with wavefront control in Section~\ref{subsec:falcosims} within an order of magnitude. These results can be compared to those reported by Galicher et al. 2020~\cite{Galicher2020} for a wrapped charge 8 vortex phase mask. Numerical simulations of the wrapped vortex phase mask predict similar performance with a 3$\sigma$ detection limit at 3.1e-8 for a dark hole from 6-13 $\lambda/D$ for narrowband light at a 10\% wavelength offset. The differences are related not just to the focal plane mask design, but also the different scoring regions, which correspond to the different inner working angle of each of these masks.

These preliminary EFC tests have shown that wrapped staircase SVCs have promise for broadband starlight suppression. The two primary reasons the SVC performs worse than the VVC are chromatic leakage and model mismatch with EFC. Even though we are able to generate a model of the wrapped staircase SVC, EFC is sensitive to any mismatch, including clocking angle, transmission uniformity, and phase steps manufacturing accuracy.

Instead of using a perfect vortex phase mask model for EFC, a more accurate model describing the steps and sectors of the wrapped staircase mask in the EFC algorithm might result in improved contrast levels and perhaps a faster convergence rate. Alternatively, a model-free wavefront control algorithm, like iEFC, might prove to be a more optimal approach for the wrapped staircase mask and other SVC topographies with sharp features\cite{Haffert2022}.


\section{Conclusion and Future Directions}

We have demonstrated the starlight suppression capabilities of a charge 6 wrapped staircase scalar vortex coronagraph in lab. We achieved average contrasts of $3.2 \times 10^{-8}$ from 3 to 10 $\lambda/D$ in monochromatic light with room for improvement by better matching the EFC model to the physical FPM. Preliminary wavefront control with EFC yielded average contrasts of $2.2 \times 10^{-7}$ from 3 to 10 $\lambda/D$ for light across a $10\%$ bandwidth. One of the drawbacks we encountered to the wrapped staircase mask design was the non-uniform transmission in the event that the star or planet lands on a non-2$\pi$ transition region between steps for an off-axis measurement. 

Although the contrasts achieved by the wrapped staircase SVC experimentally are not yet at the required level for space missions seeking exo-Earths around Sun-like stars, they are suitable for ground-based Earth-like planets around M type stars such as with the Extremely Large Telescope (ELT)\cite{Fitzgerald2022,Males2022,Kasper2021}. This study presents experimental evidence that wrapped staircase scalar vortex coronagraphs have further potential for narrowband starlight suppression, show promise in broadband light from lab results on HCST and IACT, and are a worthy avenue of further research for high contrast exoplanet imaging. These results motivate testing of a wrapped staircase scalar vortex phase mask with alternative wavefront sensing and control algorithms--particularly model-free methods~\cite{Haffert2022}. This research also yielded an improved understanding of the modal decomposition representation of a phase mask which can be used to reverse-engineer a new mask design. This can be done by selecting certain peaks to suppress: for instance the zeroth mode, odd modes, or small even modes, and then mitigating their modal intensity. Then an inverse Fourier transform is applied on this distribution to result in a new phase profile.

We plan to consider a wide range of designs including superpositions of vortices of different orders, sinusoidal azimuthal profiles, metasurface designs with nanoposts, as well as masks that include both radial and azimuthal variations. New designs will first be simulated in FALCO and then experimentally tested in lab on IACT or HCST. Further development of coronagraph design and analyses tools is required to produce phase profiles that are unequivocally more optimal than the phase ramps presented here. Furthermore, manufacturing constraints and alignment tolerances, have yet to be accounted for in the design optimization process. Moving forward for new designs, we intend to consider masks for fabrication with smooth ramps instead of staircase patterns and are also investigating various fabrication methods other than electron-beam lithography to achieve this including ultrafast laser inscription. Through further iterative design, simulation, fabrication and testing, we will continue to develop a better understanding of the capabilities and limitations of SVCs and ultimately their potential for direct exoplanet imaging.





\subsection{Disclosure}
This paper is an extension of the combined work submitted in the SPIE Proceedings: Desai et al. 2021 and Desai et al. 2022.

\appendix    


\bibliography{report}   

\begin{thebibliography}{10}

\bibitem{Akenson2013}
R.~L. {Akeson}, X.~{Chen}, D.~{Ciardi}, {\em et~al.}, ``{The NASA Exoplanet
  Archive: Data and Tools for Exoplanet Research},'' {\em Publ. Astron. Soc.
  Pac.} {\bf 125}, 989  (2013).

\bibitem{Astro2020_Report}
{National Academies of Sciences, Engineering,} and Medicine, {\em Pathways to
  Discovery in Astronomy and Astrophysics for the 2020s}  (2021).

\bibitem{HabExReport}
{The HabEx Team}, ``{The Habitable Exoplanet Observatory (HabEx) Mission
  Concept Study Final Report},'' {\em arXiv e-prints} , arXiv:2001.06683
  (2020).

\bibitem{LUVOIRReport}
{The LUVOIR Team}, ``{The LUVOIR Mission Concept Study Final Report},'' {\em
  arXiv e-prints} , arXiv:1912.06219  (2019).

\bibitem{Mawet2005}
D.~{Mawet}, P.~{Riaud}, O.~{Absil}, {\em et~al.}, ``{Annular Groove Phase Mask
  Coronagraph},'' {\em The Astrophysical Journal} {\bf 633}, 1191--1200
  (2005).

\bibitem{Foo2005}
G.~{Foo}, D.~M. {Palacios}, and J.~{Swartzlander}, Grover~A., ``{Optical vortex
  coronagraph},'' {\em Optics Letters} {\bf 30}, 3308--3310  (2005).

\bibitem{Swartzlander2005}
G.~A. Swartzlander, ``Broadband nulling of a vortex phase mask.,'' {\em Optics
  letters} {\bf 30 21}, 2876--8  (2005).

\bibitem{Lee2006}
J.~H. Lee, G.~Foo, E.~G. Johnson, {\em et~al.}, ``Experimental verification of
  an optical vortex coronagraph,'' {\em Phys. Rev. Lett.} {\bf 97}, 053901
  (2006).

\bibitem{Swartzlander2008}
J.~{Swartzlander}, Grover~A., E.~L. {Ford}, R.~S. {Abdul-Malik}, {\em et~al.},
  ``{Astronomical demonstration of an optical vortex coronagraph},'' {\em
  Optics Express} {\bf 16}, 10200  (2008).

\bibitem{Mawet2009}
D.~Mawet, E.~Serabyn, K.~M. Liewer, {\em et~al.}, ``The vector vortex
  coronagraph: Laboratory results and first light at palomar observatory,''
  {\em The Astrophysical Journal} {\bf 709}, 53 -- 57  (2009).

\bibitem{Serabyn2017}
E.~{Serabyn}, E.~{Huby}, K.~{Matthews}, {\em et~al.}, ``{The W. M. Keck
  Observatory Infrared Vortex Coronagraph and a First Image of HIP 79124 B},''
  {\em The Astrophysical Journal} {\bf 153}, 43  (2017).

\bibitem{Wang2020}
J.~J. {Wang}, S.~{Ginzburg}, B.~{Ren}, {\em et~al.}, ``{Keck/NIRC2 L'-band
  Imaging of Jovian-mass Accreting Protoplanets around PDS 70},'' {\em The
  Astronomical Journal} {\bf 159}, 263  (2020).

\bibitem{Kuhn2018}
J.~Kühn, E.~Serabyn, J.~Lozi, {\em et~al.}, ``An h-band vector vortex
  coronagraph for the subaru coronagraphic extreme adaptive optics system,''
  {\em Publications of the Astronomical Society of the Pacific} {\bf 130},
  035001  (2018).

\bibitem{Wagner2021}
K.~{Wagner}, S.~{Ertel}, J.~{Stone}, {\em et~al.}, ``{Imaging low-mass planets
  within the habitable zones of nearby stars with ground-based mid-infrared
  imaging},'' {\em arXiv e-prints} , arXiv:2107.14378  (2021).

\bibitem{Serabyn2019}
E.~{Serabyn}, C.~M. {Prada}, P.~{Chen}, {\em et~al.}, ``{Vector vortex
  coronagraphy for exoplanet detection with spatially variant diffractive
  waveplates},'' {\em Journal of the Optical Society of America B Optical
  Physics} {\bf 36}, D13  (2019).

\bibitem{Ruane2020}
G.~{Ruane}, E.~{Serabyn}, C.~{Mejia Prada}, {\em et~al.}, ``{Experimental
  analysis of the achromatic performance of a vector vortex coronagraph},'' in
  {\em Society of Photo-Optical Instrumentation Engineers (SPIE) Conference
  Series},  {\em Society of Photo-Optical Instrumentation Engineers (SPIE)
  Conference Series} {\bf 11443}, 114432O  (2020).

\bibitem{Ruane2022}
G.~{Ruane}, A.~J.~E. {Riggs}, E.~{Serabyn}, {\em et~al.}, ``{Broadband Vector
  Vortex Coronagraph Testing at NASA’s High Contrast Imaging Testbed
  Facility},'' in {\em Society of Photo-Optical Instrumentation Engineers
  (SPIE) Conference Series},  {\em Society of Photo-Optical Instrumentation
  Engineers (SPIE) Conference Series}  (2022).

\bibitem{Konig2022}
L.~{K{\"o}nig}, O.~{Absil}, M.~{Lobet}, {\em et~al.}, ``{Optimal design of the
  annular groove phase mask central region},'' {\em Optics Express} {\bf 30},
  27048  (2022).

\bibitem{Swartzlander2006}
J.~{Swartzlander}, Grover~A., ``{Achromatic optical vortex lens},'' {\em Optics
  Letters} {\bf 31}, 2042--2044  (2006).

\bibitem{Ruane2019}
G.~{Ruane}, D.~{Mawet}, A.~J.~E. {Riggs}, {\em et~al.}, ``{Scalar vortex
  coronagraph mask design and predicted performance},'' in {\em Society of
  Photo-Optical Instrumentation Engineers (SPIE) Conference Series},  {\em
  Society of Photo-Optical Instrumentation Engineers (SPIE) Conference Series}
  {\bf 11117}, 111171F  (2019).

\bibitem{Galicher2020}
R.~Galicher, E.~Huby, P.~Baudoz, {\em et~al.}, ``A family of phase masks for
  broadband coronagraphy example of the wrapped vortex phase mask theory and
  laboratory demonstration,'' {\em Astronomy {\&} Astrophysics} {\bf 635}, A11
  (2020).

\bibitem{Desai2021}
N.~Desai, J.~Llop-Sayson, N.~Jovanovic, {\em et~al.}, ``High contrast
  demonstrations of novel scalar vortex coronagraph designs at the high
  contrast spectroscopy testbed,'' in {\em Techniques and Instrumentation for
  Detection of Exoplanets X},  S.~B. Shaklan and G.~J. Ruane, Eds., {SPIE}
  (2021).

\bibitem{Mawet2009LiquidCrystal}
D.~Mawet, E.~Serabyn, K.~Liewer, {\em et~al.}, ``Optical vectorial vortex
  coronagraphs using liquid crystal polymers: theory, manufacturing and
  laboratory demonstration,'' {\em Optics Express} {\bf 17}, 1902  (2009).

\bibitem{Ganic2002}
D.~Ganic, X.~Gan, M.~Gu, {\em et~al.}, ``Generation of doughnut laser beams by
  use of a liquid-crystal cell with a conversion efficiency near 100\%,'' {\em
  Opt. Lett.} {\bf 27}, 1351--1353  (2002).

\bibitem{Mawet2005Subwavelength}
D.~Mawet, P.~Riaud, J.~Surdej, {\em et~al.}, ``Subwavelength surface-relief
  gratings for stellar coronagraphy.,'' {\em Applied optics} {\bf 44 34},
  7313--21  (2005).

\bibitem{Murakami2013}
N.~Murakami, S.~Hamaguchi, M.~Sakamoto, {\em et~al.}, ``Design and laboratory
  demonstration of an achromatic vector vortex coronagraph,'' {\em Optics
  express} {\bf 21}, 7400--10  (2013).

\bibitem{Genevet2012}
P.~Genevet, N.~Yu, F.~Aieta, {\em et~al.}, ``Ultra-thin plasmonic optical
  vortex plate based on phase discontinuities,'' {\em Applied Physics Letters}
  {\bf 100}, 013101  (2012).

\bibitem{Desai2022}
N.~Desai, J.~Llop-Sayson, A.~Bertrou-Cantou, {\em et~al.}, ``{Topological
  designs for scalar vortex coronagraphs},'' in {\em Space Telescopes and
  Instrumentation 2022: Optical, Infrared, and Millimeter Wave},  L.~E. Coyle,
  S.~Matsuura, and M.~D. Perrin, Eds.,  {\bf 12180}, 121805H, International
  Society for Optics and Photonics, SPIE  (2022).

\bibitem{Riggs2018}
A.~J.~E. {Riggs}, G.~{Ruane}, E.~{Sidick}, {\em et~al.}, ``{Fast linearized
  coronagraph optimizer (FALCO) I: a software toolbox for rapid coronagraphic
  design and wavefront correction},'' in {\em Space Telescopes and
  Instrumentation 2018: Optical, Infrared, and Millimeter Wave},  M.~{Lystrup},
  H.~A. {MacEwen}, G.~G. {Fazio}, {\em et~al.}, Eds., {\em Society of
  Photo-Optical Instrumentation Engineers (SPIE) Conference Series} {\bf
  10698}, 106982V  (2018).

\bibitem{Give'on2007}
A.~{Give'on}, B.~{Kern}, S.~{Shaklan}, {\em et~al.}, ``{Electric Field
  Conjugation - A Broadband Wavefront Correction Algorithm For High-contrast
  Imaging Systems},'' in {\em American Astronomical Society Meeting Abstracts},
   {\em American Astronomical Society Meeting Abstracts} {\bf 211}, 135.20
  (2007).

\bibitem{Ruane2018}
G.~Ruane, D.~Mawet, B.~Mennesson, {\em et~al.}, ``Vortex coronagraphs for the
  habitable exoplanet imaging mission concept: theoretical performance and
  telescope requirements,'' {\em Journal of Astronomical Telescopes,
  Instruments, and Systems} {\bf 4}, 1  (2018).

\bibitem{Llop-Sayson_SPIE2020}
J.~{Llop-Sayson}, N.~{Jovanovic}, G.~{Morrissey}, {\em et~al.}, ``{Wavefront
  control experiments with a single mode fiber at the High-Contrast
  Spectroscopy Testbed for Segmented Telescopes (HCST)},'' in {\em Society of
  Photo-Optical Instrumentation Engineers (SPIE) Conference Series},  {\em
  Society of Photo-Optical Instrumentation Engineers (SPIE) Conference Series}
  {\bf 11443}, 114432Q  (2020).

\bibitem{AVC2020}
J.~{Llop-Sayson}, G.~{Ruane}, D.~{Mawet}, {\em et~al.}, ``{High-contrast
  Demonstration of an Apodized Vortex Coronagraph},'' {\em The Astrophysical
  Journal} {\bf 159}, 79  (2020).

\bibitem{Baxter2021}
W.~Baxter, A.~Potier, G.~Ruane, {\em et~al.}, ``{Design and commissioning of an
  in-air coronagraph testbed in the HCIT facility at NASA’s Jet Propulsion
  Laboratory},'' in {\em Techniques and Instrumentation for Detection of
  Exoplanets X},  S.~B. Shaklan and G.~J. Ruane, Eds.,  {\bf 11823}, 118231S,
  International Society for Optics and Photonics, SPIE  (2021).

\bibitem{Haffert2022}
S.~Y. Haffert, J.~R. Males, K.~Van~Gorkom, {\em et~al.}, ``Advanced wavefront
  sensing and control demonstration with magao-x,''  (2022).

\bibitem{Fitzgerald2022}
M.~P. {Fitzgerald}, S.~{Sallum}, M.~A. {Millar-Blanchaer}, {\em et~al.}, ``The
  planetary systems imager for tmt: overview and status,'' in {\em Ground-based
  and Airborne Instrumentation for Astronomy IX},  C.~J. {Evans}, J.~J.
  {Bryant}, and K.~{Motohara}, Eds., {\em Society of Photo-Optical
  Instrumentation Engineers (SPIE) Conference Series} {\bf 12184}, 1218426
  (2022).

\bibitem{Males2022}
J.~R. {Males}, L.~M. {Close}, S.~Y. {Haffert}, {\em et~al.}, ``The conceptual
  design of gmagao-x: visible wavelength high contrast imaging with gmt,'' in
  {\em Adaptive Optics Systems VIII},  L.~{Schreiber}, D.~{Schmidt}, and
  E.~{Vernet}, Eds., {\em Society of Photo-Optical Instrumentation Engineers
  (SPIE) Conference Series} {\bf 12185}, 121854J  (2022).

\bibitem{Kasper2021}
M.~{Kasper}, N.~{Cerpa Urra}, P.~{Pathak}, {\em et~al.}, ``Pcs {\textemdash} a
  roadmap for exoearth imaging with the elt,'' {\em The Messenger} {\bf 182},
  38--43  (2021).

\end{thebibliography}
\bibliographystyle{spiejour}   





\end{document}